\documentclass[letter,english]{article}
\usepackage{emulateapj,onecolfloat,epsfig}

\begin{document}
\newcommand {\ds}{\displaystyle}
\newcommand {\sss}{\scriptscriptstyle}
\def\sun{\hbox{$\odot$}}

\twocolumn %
[ %
\title{Redshift Accuracy Requirements for Future 
Supernova and Number Count Surveys}

\author{Dragan Huterer$^1$, Alex Kim$^2$, Lawrence M. Krauss$^{1,3}$
and Tamara Broderick$^4$} \affil{$^1$Department of Physics and Center
for Education and Research in Cosmology and Astrophysics, Case Western
Reserve University, Cleveland, OH~~44106} \affil{$^2$Physical Sciences
Division, Lawrence Berkeley National Laboratory, Berkeley, CA~~94720}
\affil{$^3$Department of Astronomy, Case Western Reserve University,
Cleveland, OH~~44106} \affil{$^4$Laurel High School, Cleveland, OH
44106}

\begin{abstract}
We investigate the required redshift accuracy of type Ia supernova
and cluster number-count surveys in order for the redshift
uncertainties not to contribute appreciably to the dark energy
parameter error budget.  For the SNAP supernova experiment, we find
that, without the assistance of ground-based measurements, individual
supernova redshifts would need to be determined to about 0.002 or
better, which is a challenging but feasible requirement for a
low-resolution spectrograph. However, we find that accurate redshifts
for $z<0.1$ supernovae, obtained with ground-based experiments, are
sufficient to immunize the results against even relatively large
redshift errors at high $z$. For the future cluster number-count surveys
such as the South Pole Telescope, Planck or DUET, we find that the
purely statistical error in photometric redshift is less important,
and that the irreducible, systematic bias in redshift drives the
requirements. The redshift bias will have to be kept below 0.001-0.005
per redshift bin (which is determined by the filter set), depending on
the sky coverage and details of the definition of the minimal mass of
the survey.  Furthermore, we find that X-ray surveys have a more
stringent required redshift accuracy than Sunyaev-Zeldovich (SZ)
effect surveys since they use a shorter lever arm in redshift;
conversely, SZ surveys benefit from their high redshift reach only so
long as some redshift information is available for distant ($z\gtrsim
1$) clusters.
\end{abstract}

 ]

\section{Introduction}

Two of the most promising methods to measure cosmological parameters,
in particular those describing dark energy, are distance measurements
of type Ia supernovae (SNe Ia) and number counts of clusters of
galaxies in the universe. SNe Ia have provided original direct
evidence for dark energy (Riess et al.\ 1998, Perlmutter et al.\ 1999)
(for earlier, indirect evidence, see Krauss and Turner\ 1995 or
Ostriker and Steinhardt\ 1995) and are currently the strongest direct
probe of the expansion history of the universe (Tonry et al.\ 2003,
Knop et al.\ 2003). Their principal strength is the simplicity of
relating the observable -- which is essentially the luminosity
distance -- to cosmological parameters, and also the fact that each
supernova redshift-magnitude pair provides a distinct measurement of a
combination of those parameters. Number-counts, on the other hand, use
the fact that galaxy clusters are the largest collapsed structures in
the universe that have undergone a relatively small amount of
post-processing.  Their distribution in redshift can be reliably
calculated in a given cosmological model. The evolution of cluster
abundance is principally sensitive to the comoving volume and growth
of density perturbations (Haiman, Mohr \& Holder 2001) and this
cosmological probe is expected to reach its full potential with
upcoming and future wide-field surveys.

Rapid improvement in the accuracy of measuring cosmological parameters
implies that various systematic uncertainties, previously ignored, now
have to be controlled and understood quantitatively. In the case of
supernova measurements, an example is provided by the proposed
SuperNova/Acceleration Probe (SNAP) satellite (Akerlof et al.\ 2004)
whose goals for measuring the equation of state of dark energy $w$ and
its variation with redshift $dw/dz$ drive the requirements on the
systematic control that are considerably more stringent than those
attainable with current surveys. Similarly, the principal systematic
difficulty in cluster counts is in establishing the relation between
observable quantities (X-ray temperature or Sunyaev-Zeldovich flux
decrement), and the cluster's mass which is necessary for comparison
with theory. The mass-temperature relation, for example, is known to
have a considerable scatter and is currently poorly determined, with
fairly large intrinsic statistical errors and considerable systematic
disagreements between different authors (see e.g. Fig. 2 in Huterer \&
White 2002). The cleanest way to include the mass-observable relation
might be to determine it from the survey itself (this is known as
``self-calibration''; Levine, Schulz \& White 2002, Majumdar \& Mohr
2003, Hu 2003, Lima \& Hu 2004), but this will almost certainly lead
to degradations in parameter accuracies. Future surveys will require a
careful accounting of all systematics -- theoretical and
observational.

In this paper we concentrate on one of the most basic ingredients of
supernova and cluster count measurements: the determination of
redshift.  In the case of SNe Ia spectroscopic observations are
necessary to identify the supernova type, and redshift is then
supplied for free. Recently completed and ongoing surveys have
sufficiently poor magnitude uncertainty and relatively low statistics
and relatively weak control on known systematics, so that the
spectroscopic redshift error is small enough for the redshifts to be
considered perfectly known. However, as we shall see, future supernova
observations require such accurate redshifts that even the
spectroscopic accuracy is not {\it a priori} guaranteed to be
sufficient.

In the cluster count case, the situation is even more interesting, as
spectroscopic observations will not be possible for all clusters, which
may number in the tens of thousands.  One will therefore rely on
photometric redshifts.  Although photometric redshifts are already
impressively accurate (e.g. Fern\'{a}ndez-Soto et al.\ 2002, Csabai et
al.\ 2003, Collister \& Lahav 2003, Vanzella et al.\ 2003), we shall
find that their bias (the difference between the mean photometric
value and the true value at any redshift) needs to be kept
exceedingly small for the redshift error not to contribute appreciably
to the total error budget. Our analysis is timely, as follow-up
surveys to obtain cluster redshifts, such as that at Cerro-Tololo
International Observatory, are about to get underway soon.  Our
analysis also complements recent analysis of the effect of systematic
errors on future SN Ia measurements (Kim et al.\ 2003, Frieman et
al.\ 2003) and a variety of related analyses regarding the cluster
number-count surveys (e.g. Bartlett 2000, Holder \& Carlstrom 2001,
White, Hernquist \& Springel 2002, White, van Waerbeke \& Mackey 2002,
Benson, Reichardt \& Kamionkowski 2002, White 2003).

The paper is organized as follows. In section~\ref{sec:method} we
outline the procedure to include the redshift uncertainty in the
standard Fisher-matrix parameter estimation. In Sec.~\ref{sec:sne} we
discuss the redshift requirements for future supernova surveys, while
in Sec.~\ref{sec:counts} we do the same for future cluster count
surveys. We conclude in Sec.~\ref{sec:concl}. Our fiducial model is a
flat universe with matter energy density relative to critical of
$\Omega_M=0.3$ and the equation of state of dark energy $w=-1$.  Other
cosmological parameters, necessary for the cluster abundance
calculation, are discussed in Sec.~\ref{sec:counts}.

\section{Methodology}
\label{sec:method}
Let us assume we have an observable $O(z)$ from which we want to
determine $P$ cosmological parameters $\theta_1, \ldots,
\theta_P$. Since the number of observed objects is large (thousands
for future SNe Ia surveys and of order ten thousand for cluster
surveys), we bin observations in $Q$ redshift bins centered at $z_k$
($1\leq k\leq Q$) each with width $\Delta z_k$.

To include the redshift uncertainty, we treat the bin centers $z_1,
\ldots, z_Q$ as $Q$ additional nuisance parameters
$\theta_{P+1},\ldots, \theta_{P+Q}$. Variation of any of these
parameters moves the location of the whole corresponding redshift bin.
We use the Fisher matrix formalism in order to estimate all $P+Q$
parameters simultaneously, and we give priors to redshift parameters
that represent how accurately they are independently determined. 

The $(P+Q)\times (P+Q)$ Fisher matrix is given by

\begin{equation}
F_{ij}=\sum_{k=1}^{Q} {N_k\over \sigma_O(z_k)^2}
{\partial O(z_k)\over \partial \theta_i}
{\partial O(z_k)\over \partial \theta_j}
\label{eq:fisher}
\end{equation}

\noindent where $N_k$ is the number of objects in the k-th bin.  The
observable $O(z_k)$ is the mean apparent magnitude of a supernova $m(z_k)$ in
a given redshift bin, or else the number of clusters $N(z_k)$ in a bin.
Note that the redshift parameter representing the $i$th bin affects
the observable $O(z_k)$ only if $i=k$. Therefore

\begin{equation}
 {\partial O(z_k)\over \partial \theta_{P+i}}=
 {\partial O(z_i)\over \partial \theta_{P+i}}\,\delta_{ik} \quad{\rm for}\quad 
   i\in  [1, \ldots, Q]
\end{equation}

\noindent and the expression for the Fisher matrix simplifies
accordingly for the redshift parameter terms.

We assume the error in the observable $O(z_k)$ is Gaussian-distributed with
standard deviation $\sigma_O(z_k)$. With that assumption, any Gaussian
prior imposed on the individual object, $\sigma_{\rm prior}$ will be
equivalent to imposing a prior of $\sigma_{\rm prior}/\sqrt{N_k}$ on
the observable representing the k-th redshift bin since there are
$N_k$ objects in this bin.  Note that the number of redshift bins,
$Q$, needs to be large enough to retain the shape information of the
function $O(z)$; we use steps of 0.02 for both SNe Ia and number
counts, and have checked that a higher number of bins leads to
negligible changes in all of our results. Note too that these bins are
used for computational accuracy; they should not be confused with the
{\it physical} redshift bins which are determined by the filter set of
the experiment and which we later discuss.  Finally, we ignore the
cosmic variance contribution to the error in number-count surveys,
since it has been shown that cosmic variance becomes small for the
high-redshift cluster surveys with relatively high mass threshold (Hu
\& Kravtsov 2003), which is the case we study in this paper.

While our Fisher matrix formalism assumes the redshift errors to be
Gaussian, it is conceivable that the errors will have significant
non-Gaussian tails and/or pronounced skewness. This may especially be
true for the photometric redshift errors, where a small fraction of
redshifts may have a large deviation from their true value.  In this
situation our formalism is still appropriate: by the Central Limit
Theorem (and as confirmed with Monte Carlo) the central (mean) value
of any given redshift bin, $z_i$, is guaranteed to have Gaussian error
dispersion (whose center may be shifted from the true value of $z_i$)
{\it even if individual objects in this bin have errors that are
strongly non-Gaussian}. This fully justifies our use of Gaussian priors
on $z_i$, and the same argument applies to other observables we
consider --- mean supernova magnitude and number of clusters in a
redshift bin. In order to {\it determine} the resultant RMS of $z_i$
and its bias (shift relative to its true value), however, one needs to
know {\it a priori} the redshift distribution of
objects and their measurement errors. While detailed modeling
of the photometric redshift error (taking into account various types
of galaxies and their properties at any given redshift) will require a
Monte Carlo approach that is outside the scope of this work, here we
explore results for a range of widths of the Gaussian distribution of
the redshift bin centers $z_i$.

\section{Type Ia supernovae}
\label{sec:sne}

\subsection{Redshift dependence of supernova measurements}

The measurement of the cosmological parameters using calibrated candles
requires both the magnitude and redshift of the object in question.
In supernova studies the redshift measurement is typically taken from
the spectrum of the host galaxy, either from sharp emission lines or
from the 4000 \AA\ break.  Up to the present, the magnitude error of
high-redshift supernovae have dwarfed the redshift measurement error.  As we
enter an era of high-precision supernova cosmology, with significant
improvement in statistical and systematic uncertainties, we need to
explore the effects of redshift measurement error on the determination
of cosmological parameters.

Cosmological observations of high-redshift supernovae are generally
made in observer $X$ and $Y$ bands which roughly correspond
to supernova-frame $B$ and $V$ bands.
Observed Type Ia supernova magnitudes are modeled as
\begin{eqnarray}
m_X &= &M_B  - \alpha(s-1) + A_B(s,z) +K_{BX}(s,z)+   \nonumber\\[0.07cm] 
&&\mu(z;\theta_i).
\label{distmod:eqn}
\end{eqnarray}
The peak absolute magnitude of a supernova $M_B - \alpha(s-1)$ is a
function of the ``stretch'' $s$ of its light-curve shape; supernovae
with higher stretch are intrinsically brighter (Perlmutter  et
al.\ 1997).  The K-correction $K_{BX}$, a function of redshift and
stretch, accounts for differences in the spectral energy distribution
(SED) transmitted through the $B$ and $X$ bands for low- and
high-redshift supernovae respectively.  The extinction from the host
galaxy is given by $A_B=R_BE(B-V)$ where the color excess is
\begin{eqnarray}
E(B-V)&= & \left [(m_X-K_{BX}(s,z))- (m_Y-K_{VY}(s,z))\right ]
\nonumber\\[0.1cm]
&-&  (B-V)_{\rm expected}(s)
\end{eqnarray}
where $(B-V)_{\rm expected}(s)$ is the expected supernova color, which
is a function of stretch. The distance modulus $\mu$ is a function of
redshift and the cosmological parameters.  Gravitational lensing
magnification is not considered here since its effect on the inferred
magnitudes of distant supernovae is not sensitive to small variations
in redshift.

The effect of redshift error on the estimated distance modulus is
straightforward: a positive redshift error, $dz$, incorrectly gives an
inflated distance to the supernova.  In addition, the measurement of
stretch is itself dependent on $z$; the stretch is obtained from the
observed width $W$ of a light curve using the formula $s=W/(1+z)$ so
that an error in $z$ propagates as $ds=-s\,dz/(1+z)$.  An overestimate
of redshift gives an underestimate of the stretch factor and therefore an
overestimate of the expected supernova magnitude.

The extinction terms and K-correction depend on both redshift and
stretch.  Expected observed colors with a fixed pair of
filters near the restframe $B$ and $V$ are bluer for slightly higher
redshift supernovae.  An overestimated redshift will thus give an
overestimated extinction determination.  In contrast, the simultaneously
underestimated stretch determination overestimates the intrinsic
redness of the supernova, underestimating the extinction.
Additionally, stretch-dependent SED's and redshift errors introduce
K-correction errors whose behavior depends on the specific redshift
and filters involved.

We propagate redshift errors into errors in the expected observed peak
magnitude for a canonical supernova search.  We adopt a filter-set
consisting of redshifted $B$ filters such that the $n$-th filter has
throughput $F_n(\lambda)=B(\lambda/1.16^n)$, where $\lambda$ is
wavelength.  We adopt the empirically derived $\alpha=1.9$
stretch--magnitude relation found by Perlmutter et al.\ (1997).  The
stretch-dependent supernova color is given as $B-V = -0.19(s-1) -
0.05$.  We use the K-correction methodology given in Nugent, Kim, \&
Perlmutter (2002); the observer filters with effective wavelengths
closest to $4400(1+z)$ and $5500(1+z)$ are associated with rest-frame
$B$ and $V$ respectively.  The host-galaxy extinction is assumed to
obey the standard $R_B=4.1$ dust model of Cardelli, Clayton, \& Mathis
(1989).

\begin{figure}[!t]
\vspace{0.5cm}
\epsfig{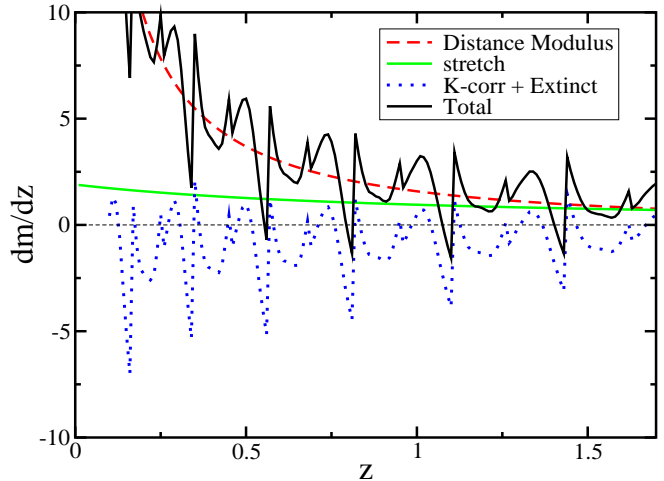}
\caption{The differential effect on the expected supernova peak magnitude
with variation in redshift, or $dm/dz$, as a function of redshift.
Also shown are the contribution from individual terms;
the distance modulus, stretch, K-correction, and
extinction. }
\label{fig:dmdz}
\end{figure}

Figure~\ref{fig:dmdz} shows individual contributions to the derivative
of magnitude with respect to redshift, $dm/dz$, as well as their sum
for an $s=1$ supernova.  The K-correction and extinction errors are
discontinuous and periodic in redshift as different observer filters
are traversed. The distance modulus error, being proportional to the
relative error in luminosity distance, $d(d_L)/d_L$, goes to infinity
as redshift goes to zero. This is simple to understand, as in the
$z\rightarrow 0$ limit $d(d_L)/dz$ approaches a constant, while $d_L$
itself goes to zero. Therefore, low-$z$ supernovae have the most need
for accurate redshifts; we discuss this further below. Note that the
use of $dm/dz$ to assess the effect of redshift errors is only
approximate since the K-corrections can be highly nonlinear; however,
the approximation is good for the span of redshift errors that we
consider.

\subsection{Results}

We explore the effects of redshift error in the measurement of dark
energy parameters based on a supernova search such as the proposed
SNAP space telescope (Akerlof et al.\ 2004). As we discuss below,
more powerful experiments require more stringent control of redshift
errors; therefore, our requirements for SNAP will be more than
sufficient for other ground and space-based surveys in the next 5-10
years. We assume a supernova distribution with around 2700 SNe
distributed between $z=0.1$ and $1.7$ together with 300 additional
low-$z$ ($z\approx 0.05$) SNe from the ground-based SN Factory
(Aldering et al.\ 2002). The fiducial magnitude error per
supernova, the quadratic sum of measurement error and intrinsic
supernova magnitude dispersion, is 0.15 magnitudes. (An analysis of
the effect of redshift-dependent magnitude uncertainties will be
discussed in Krauss et al., in preparation.) We consider the
degradation, due to imperfectly known redshifts, of the accuracy in
the equation of state of dark energy $\sigma_w$, where $w$ is assumed
constant. The uncertainty $\sigma_w$ is computed by marginalizing over
the matter density $\Omega_M$, the overall offset in the
magnitude-redshift diagram, $\mathcal{M}$, and the redshift parameters
$z_i$ ($i\in [1, \ldots, Q]$). We also consider the degradation in the
accuracy in measuring the redshift evolution in the equation of state,
$dw/dz$, where $w(z)=w_0+z\,dw/dz$; in this case we further
marginalize over $w_0$ and add a Gaussian prior of $0.01$ to
$\Omega_M$, to allow comparisons with other analyses in which this
procedure has become standard.

\begin{figure}[!t]
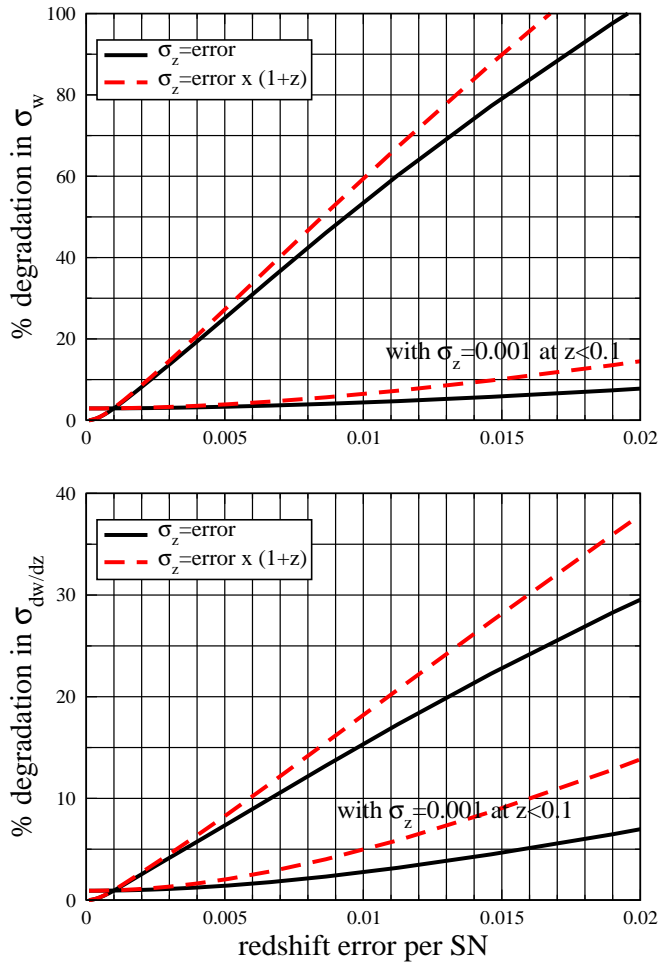

\epsfig{file=degrade_w.eps, height=2.5in, width=3.4in}\\
\epsfig{file=degrade_dwdz.eps, height=2.5in, width=3.4in}
\caption
{Degradation in the measurement accuracy of the equation of state $w$
(top panel) and, alternatively, its rate of change with redshift
$dw/dz$ (bottom panel) as a function of redshift error per supernova. The
solid line shows the case when the redshift error is constant for all
SNe, while the dashed line shows the case when the error (plotted on
the abscissa) is multiplied by (1+z). The flatter pair of curves in
each panel corresponds to the case when $z<0.1$ SNe were assumed to
have a fixed redshift error of 0.001 per SN.}
\label{fig:degrade_w}
\end{figure}

Figure~\ref{fig:degrade_w} shows the degradation in $\sigma_w$ (top
panel) and $dw/dz$ (bottom panel) as a function of the redshift error
per supernova.  The solid line shows the case when the redshift error
is constant for all SNe, whereas the dashed line represents an
uncertainty growing as $dz\propto (1+z)$.  The two cases are
qualitatively similar, and show that, for example, a redshift error of
0.005 per SN leads to a 25\% increase in $\sigma_w$ and a 7\%
increase in $\sigma_{dw/dz}$.

However, these results assume that the error in redshift, absolute or
fractional, is the same at all redshifts.  In reality, the SN Factory
spectra will have a fixed high resolution. The flatter
pairs of lines in Fig.~\ref{fig:degrade_w} assume that redshifts at
$z<0.1$ have a fixed accuracy of 0.001 per SN, while those at $z>0.1$
have the accuracy shown. The degradation in $w$ or $dw/dz$ is now
smaller than about 10\%, even for errors of 0.02 for SNe at $z>0.1$!
Therefore, accurate redshifts of low-$z$ supernovae immunize against
larger errors at higher redshifts. This conclusion is easy to
understand: fixed error in redshift roughly corresponds to fixed error
in distance to a supernova, while the total SN magnitude error
increases linearly with distance. Therefore, the redshift error
contributes a larger percent of the total error budget for
low-redshift supernovae. Furthermore, low-$z$ supernovae are crucial for
parameter determination and their omission (or inclusion, but with
large redshift error) would lead to nearly a factor of two
degradation in the constraints on $w$ and $dw/dz$ (e.g. Huterer \&
Turner 2001).  Therefore, we conclude that, provided that redshifts of
low-$z$ supernovae are measured with high accuracy, measurements of $w$
and $dw/dz$ are weakly sensitive to the redshift errors of high-$z$ SNe.

\begin{figure}[!t]
\epsfig{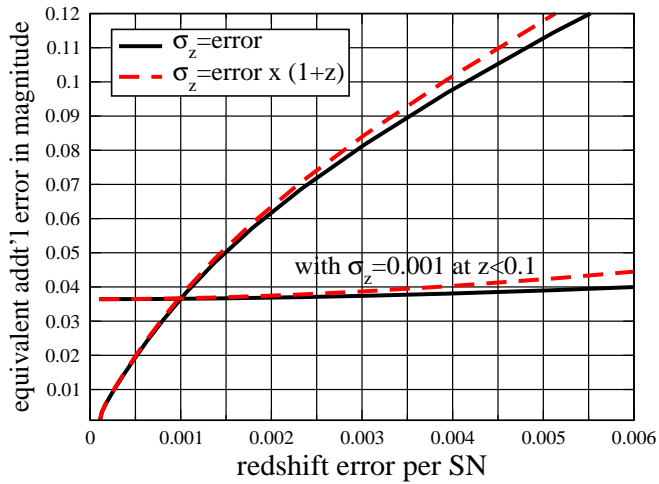}
\caption
{Redshift error increases the degradation in $\sigma_w$ as some
equivalent {\it additional} magnitude error, which is shown on the
ordinate. The solid line shows the case when the redshift error is
constant for all SNe, while the dashed line shows the case when the
error (plotted on the abscissa) is multiplied by $(1+z)$.  As in
Fig.~\ref{fig:degrade_w}, the flatter pair of curves corresponds to
the case when $z<0.1$ SNe were assumed to have a fixed redshift error
of 0.001 per SN.}
\label{fig:zerr_magerr}
\end{figure}

For convenient reference, we associate redshift errors with the
magnitude error that gives an equivalent uncertainty in $w$; see
Fig.~\ref{fig:zerr_magerr}.
For example, a 0.005 redshift error introduces an uncertainty
equivalent to the additional 0.1 intrinsic magnitude dispersion per
SN.  As before, the flatter pair of lines in the same Figure shows the
effect of accurate redshifts for $z<0.1$ SNe, in which case the
overall redshift uncertainty contributes little ($\lesssim 0.02$ mag in
the range of redshift errors shown) to the total error budget.

The spectroscopic follow-up of high-redshift supernovae (or more
appropriately, their host galaxies) from surveys must be carefully
considered.  Ground-based spectroscopy associated with current
high-$z$ supernova searches have more than sufficient resolution to
measure redshifts to $\lambda/\delta\lambda=200$.  (Subpixel
interpolation gives a wavelength resolution several times better than
the instrumental resolution $R$.) Very wide-field supernova searches
that discover thousands or more supernovae may depend on photometric
redshifts as an alternative to spectroscopic followup.  Photometric
redshift determination is currently limited by a statistical accuracy
floor of a few percent (see Sec.~\ref{sec:counts}); in this case the
effect of redshift error can be comparable in size to the intrinsic
corrected-magnitude dispersion of SNe Ia of $\sim$0.1 mag, although we
have just shown that accurate redshifts at $z<0.1$ will largely
immunize against the overall redshift contribution to the error
budget.  Of course, the two more serious problems are identification
of SNe Ia and control of systematics, both of which are very difficult
without their spectra.

In the case of a 2-m space telescope observing $z \sim 1.7$ supernovae
(such as SNAP), the Poisson noise from the zodiacal background and
source can be low.  Considering the noise properties of HgCdTe
detectors, signal-to-noise arguments push for a low-resolution
spectrograph to avoid a detector-noise-limited instrument.  If the
on-board spectrograph is to provide supernova redshifts, the
competing needs for low and high resolution must be considered in the
design.

We end with several comments. First, we do not consider the peculiar
velocities since they have negligible effect on high-redshift
supernovae.  Second, our supernova calculations involve statistical
magnitude errors only. Adding the irreducible systematic error in each
redshift bin, as done in Frieman et al.\ (2003) for example, is
straightforward and leads to a slight weakening of the required
redshift accuracy. This is easy to understand, since the fiducial
parameter accuracy, such as $\sigma_w$, slightly weakens in the
presence of systematics, and the redshifts do not need to be known
quite as accurately as in the perfect world without systematics.

Finally, we comment on the possibility of using the {\it photometric}
redshifts for non-local ($z>0.1$) supernovae.  In this approach one
has to consider the irreducible redshift errors within coarse redshift
bins; this point is discussed in detail in \S~4.1.  In the supernova
survey considered here, and assuming correlation length of $0.15$ in
redshift, approximately 240 supernova would fall in each coarse
redshift bin.  A redshift uncertainty per supernova of $\delta z_{\rm
sn}$ can then be viewed as being equivalent to a redshift uncertainty
per bin of $\delta z_{\rm bin}=\delta z_{\rm sn}/\sqrt{240}$.  Using
this relation, Figure 2 can provide a first estimate of the
degradation in the measurement of $w$ and $dw/dz$ from these
irreducible photometric errors.  For example, a redshift bias per bin
smaller than $0.001$ would be necessary to ensure that the degradation
in the measurement of $w$ and $dw/dz$ is less that $5\%$. Clearly,
photometric redshifts would need to be exceedingly accurate in order
not to degrade the dark energy constraints from type Ia supernovae.

\section{Cluster number-count surveys}
\label{sec:counts}

\subsection{Fiducial surveys and assumptions}

Clusters can be found using their X-ray flux; through their
Sunyaev-Zeldovich temperature decrement, or through deflection of
light from background galaxies due to weak gravitational lensing by
the cluster.  Current or upcoming surveys specifically suited to find
clusters include XMM Serendipitous Cluster Survey (Romer et al.\ 2001)
XMM Large Scale Structure survey (Pierre et al.\ 2003),
ACBAR\footnote{http://cosmology.berkeley.edu/group/swlh/acbar/},
Sunyaev-Zeldovich Array\footnote{http://astro.uchicago.edu/sza/}, APEX
SZ survey\footnote{http://bolo.berkeley.edu/apexsz/} and Atacama
Cosmology
Telescope\footnote{http://www.hep.upenn.edu/~angelica/act/index.html}. Cluster
cosmology will realize its full potential with future wide-field
surveys, such as the South Pole
Telescope\footnote{http://astro.uchicago.edu/spt/} (SPT), the
space-based mission
Planck\footnote{http://astro.estec.esa.nl/Planck/}, the proposed
space-based mission DUET, and, in the next decade, SNAP.

Cluster redshifts are required in order to use clusters as a probe of
cosmology, yet the large number of clusters expected in the
aforementioned upcoming surveys (thousands to tens of thousands) makes
it impractical to obtain their redshifts spectroscopically. Therefore,
cluster abundance studies will rely on the photometric
redshifts. Fortunately, cluster photometric redshifts are currently
measured with very good accuracy (e.g.\ Bahcall et al.\ 2003) chiefly
because the photo-z's of individual galaxies in the cluster can be
averaged, leading to statistical errors in cluster redshifts of about
0.02.  However, goals for the cluster abundance surveys are set high
-- measuring the equation of state of dark energy $w$ to an accuracy
of 5-10\% and the power spectrum normalization $\sigma_8$ to about 1\%
-- and it is worthwhile to study the required photometric redshift
accuracy in order to achieve this goal. Previous studies of the
cluster abundance (Haiman, Mohr \& Holder 2001, Levine, Schulz \&
White 2002, Battye \& Weller 2003, Hu \& Kravtsov 2003, Mohr \&
Majumdar 2003a, Molnar et al.\ 2003) have explored the efficacy of
cluster number counts as a probe of cosmology, but all assumed perfect
knowledge of redshifts.

We adopt the fiducial cosmological model which is in accordance with
recent results from the WMAP experiment (Spergel et al.\ 2003) and the
Sloan Digital Sky Survey (Tegmark et al.\ 2003). We assume a flat
universe with matter energy density relative to critical
$\Omega_M=0.3$, dark energy equation of state $w=-1$, and power
spectrum normalization $\sigma_8=0.9$.  We use the spectral index and
physical matter and baryon energy densities with mean values $n=0.97$,
$\Omega_M h^2=0.140$ and $\Omega_B h^2=0.023$ respectively. We add a
fairly conservative prior of 5\% to each of these parameters; we
checked that our results are insensitive to this prior. The parameters
to which cluster surveys are most sensitive are $\Omega_M$, $w$ and
$\sigma_8$, and we do not give any priors to these parameters. The
fiducial surveys we consider determine $\Omega_M$ and $\sigma_8$ to
accuracy of about 0.01 and $w$ to about 0.02-0.12. We are, however,
only interested in the {\it degradation} of these accuracies due to
uncertain knowledge of redshifts, and this fact makes our results less
dependent on the details of the survey. Finally, for this analysis we
assume that the 'mass-observable' (i.e. mass-temperature, or
mass-X-ray flux) relation is perfectly known. We have checked that
leaving the normalization of this relation as a free parameter to be
determined by the data can strongly degrade the fiducial parameter
constraints, but it affects the sensitivity to the knowledge of
redshifts, which we explore here, much more weakly.

To compute the comoving number of clusters we use the Jenkins et al.\
(2001) mass function. The required input is the linear power spectrum;
for $w=-1$ models, we use the formulae of Eisenstein \& Hu (1997)
which were fit to the numerical data produced by CMBfast (Seljak and
Zaldarriaga 1996). We generalize the formulae to $w\neq -1$ by
appropriately modifying the growth function of density
perturbations. The total number of objects in any redshift interval
centered at $z$ and with width $\Delta z$ is

\begin{equation}
N(z, \Delta z)=\Omega_{\rm survey}
\int_{z-\Delta z/2}^{z+\Delta z/2} n(z, M_{\rm min}(z)) 
\,{dV(z)\over d\Omega\,dz}\, dz
\end{equation}

\noindent where $\Omega_{\rm survey}$ is the total solid angle covered
by the survey, $n(z, M_{\rm min})$ is the comoving density of clusters
more massive than $M_{\rm min}$, and $dV/d\Omega dz$ is the comoving
volume element. 

An incorrect determination of individual cluster redshifts will lead to
an incorrect central value of the redshift bin to which these clusters
are assigned (see below for the definition of redshift bins). This in
turn leads to evaluating the theoretically expected number of clusters
$N(z, \Delta z)$ at an incorrect central redshift $z$, thus biasing 
the inferred cosmological parameters. Here we represent the
uncertainty in the central value of the redshift bin as

\begin{equation}
\sigma_{\rm z, bin}=\sqrt{{\sigma_{\rm clus}^2\over N(z, \Delta z)} + 
\sigma_{\rm sys}^2},
\label{eq:sigma_bin}
\end {equation}

\noindent that is, the redshift error is the sum of the purely
statistical (random Gaussian) error per cluster $\sigma_{\rm clus}$
and an irreducible, systematic error $\sigma_{\rm sys}$.  The source
of the irreducible error could be, for example, a systematic offset in
the photometric error determination which affects all clusters in that
bin equally. We assume that the irreducible error is uncorrelated
between bins.  The Fisher matrix is constructed as in
Eq.~(\ref{eq:fisher}), with $O(z)\equiv N(z, \Delta z)$.

We assume three representative fiducial surveys: 1) South Pole
Telescope (SPT), a 4000 sq.\ deg.\ survey that will detect clusters
through their SZ signature, 2) Planck mission, which we consider to be
a 28,000 sq.\ deg.\ SZ survey and 3) DUET, a planned X-ray space
mission with coverage of 10,000 sq.\ deg. For the SPT and Planck we
use the mass-SZ flux relation from Majumdar \& Mohr (2003b), while for
DUET we assume the Majumdar-Mohr mass-Xray flux relation. We normalize
these analytical 'mass-observable' relations so that all three of
these surveys would produce between 18,000 and 25,000 clusters for our
fiducial cosmology; we have checked that different normalizations do
not change our results dramatically. Nevertheless, the choice of the
mass-observable relation is important since $M_{\rm min}$ depends on
cosmological parameters and modifies the error budget. To present a
range of possibilities, we further consider the SPT survey with fixed
limiting masses of $10^{14}$, $2\times 10^{14}$ and $5\times 10^{14}\,
h^{-1}M_{\sun}$; in our fiducial cosmology these three possibilities
lead to about 95,000, 20,000 and 1,700 clusters respectively.

Redshifts for the clusters are provided by optical and near-infrared
photometric follow-up. These photometric redshifts are calibrated
using supplemental spectroscopic observations of a galaxy subset. We
assume the Sloan Digital Sky Survey (SDSS) passbands; they have been
extensively used for photometric-redshift measurements of
moderate-redshift galaxy clusters (Bahcall et al.\ 2003).  The SDSS
photometric system (Fukugita et al.\ 1996) is comprised of five bands:
$u'$ peaks at 3500\AA\ with FWHM of 600\AA, $g'$ peaks at 4800\AA\
with FWHM of 1400\AA, $r'$ peaks at 6250\AA\ with FWHM of 1400\AA,
$i'$ peaks at 7700\AA\ with FWHM of 1500\AA, and $z'$ peaks at
9100\AA\ with FWHM of 1200\AA. The redshift bins are defined by where
the 4000\AA\ line enters and leaves these bands, and the bin width is
typically $0.1$-$0.2$ in redshift.  At redshifts greater than about
1.2 the 4000\AA\ line leaves the observable bands and enters the
infrared, which makes obtaining the photometric redshifts at higher $z$
much more difficult. To represent the situation in a few years, we
assume the redshift bin widths of $0.5$ at $z>1.2$, keeping the error
per bin the same; this is roughly equivalent to doubling the error per
redshift interval found at $z<1.2$. We quantitatively discuss the
redshift accuracy at high redshift in the following subsection.

\begin{figure}[!t]
\epsfig{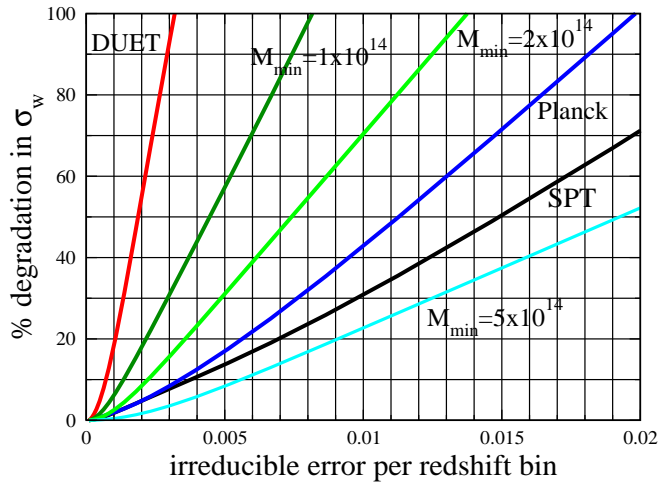}
\caption
{Degradation in the measurement accuracy of the equation of state as a
function of the {\it irreducible} redshift error per bin. SPT and
Planck curves use the mass-SZ flux relation to compute $M_{\rm min}$,
while the DUET curve uses the mass-Xray flux relation. The three other curves
show the SPT survey with fixed $M_{\rm min}$ in units of $h^{-1}M_{\sun}$.}
\label{fig:degrade_w.dndz}
\end{figure}

\subsection{Results}

The {\it purely statistical} error in photometric redshifts,
$\sigma_{\rm clus}$, is largely irrelevant for cosmological
constraints, as can be seen from Eq.~(\ref{eq:sigma_bin}), since the
large number of clusters (in all redshift bins except for those at the
highest redshifts) will make error per bin very small.  Therefore, the
current statistical error with scatter of about $\sigma_{\rm
clus}=0.02$ in redshift contributes negligibly, by itself, to the
total error budget, and we assume this statistical error for each
individual cluster.  However, we find that the results are sensitive
to {\it uncorrelated irreducible systematic} errors in redshift bins,
$\sigma_{\rm sys}$.  The measurement of photometric redshift primarily
relies on the position of the 4000\AA\ break.  For each redshift there
is a corresponding filter (or overlapping filter pair) which is
sensitive at $4000(1+z)$\AA.  Galaxies at similar redshift and whose
redshift determinations rely primarily on the same filters are thus
susceptible to common systematic errors.  These errors can arise due
to improper modeling of the filters, photometric calibration
uncertainty, or statistical errors in the redshift calibration
process.

Figure \ref{fig:degrade_w.dndz} shows the degradation in the accuracy
in $w$ as a function of the irreducible error. We show the degradation
in $w$ as a representative example, and have checked that degradations
in $\Omega_M$ and $\sigma_8$ are comparable and their range of
sensitivities is spanned by the various curves shown in this
figure. We see that the X-ray survey is most sensitive to this
redshift error; this is not surprising as the X-ray survey runs out of
clusters at $z\gtrsim 1.0$ (most of them are actually at $z\lesssim
0.6$) and hence uses a shorter lever arm in redshift than SZ surveys.
The differences in various curves show the dependence of the
sensitivity of $\sigma_w$ on details of the survey and, in particular,
of the definition of the minimal cluster mass.  The irreducible error
per bin has to be kept below 0.001-0.005, depending on the survey
details, in order for it not to contribute more than $\sim 10\%$ to
the error in $w$ and other cosmological parameters.  Furthermore, we
have explored a range of fiducial surveys, varying the parameter set,
sky coverage, and the details of the mass-observable relation, and
found that surveys with less power to measure cosmological parameters
typically have weaker requirements on the redshift accuracy. While we
have shown a range of possibilities in Fig.~\ref{fig:degrade_w.dndz},
we note that the exact requirements on the redshift accuracy for any
given survey will be known only after the survey in question has
started its operation and the accuracy of cluster mass determination
from the observed flux or temperature becomes known.  

\begin{figure}[!t]
\epsfig{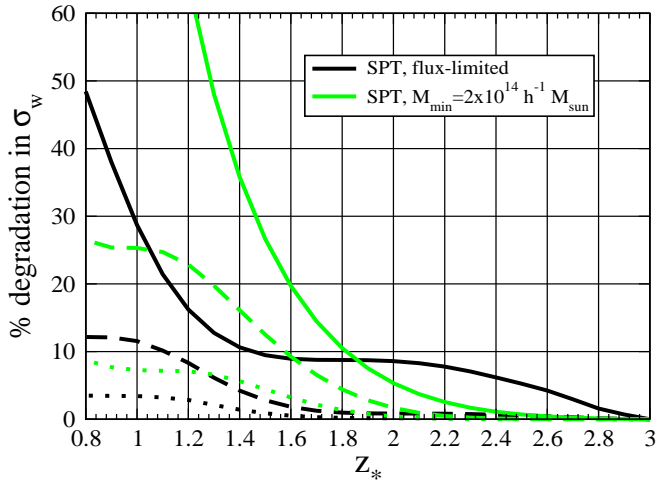}
\caption
{Degradation in the measurement accuracy of the equation of state as a
function of the redshift $z_*$ beyond which photometric redshift
information is poor or nonexistent. Redshift information is assumed to
be perfect at $z<z_*$, while at $z>z_*$ we alternatively assume: no
redshift information (solid lines), irreducible errors of 0.05 per
redshift bin width of $\Delta z=0.1$ (dashed lines) and 0.02 per
redshift bin (dotted lines). We show cases for the flux-limited SPT and
for SPT with fixed $M_{\rm min}$. Note that an X-ray survey would show
no degradation in $\sigma_w$ at all, since essentially all of its
clusters are at redshifts $\lesssim 1$. }
\label{fig:z_break}
\end{figure}

We mentioned earlier that the future accuracy of photometric redshifts
at $z\gtrsim 1$ is uncertain.  We explore the accuracy in measuring
$w$ on redshift information at $z>z_*$ where we let $z_*$ vary between
0.8 and 3.0. Figure~\ref{fig:z_break} shows that not obtaining
redshifts for clusters at such redshifts can significantly degrade the
performance of SZ surveys (X-ray survey do not have this problem,
since they have very few clusters at $z\gtrsim 1$). Redshift
information is assumed to be perfect at $z<z_*$, while at $z>z_*$ we
alternatively assume no redshift information (solid lines),
irreducible errors of 0.05 per
redshift bin width of $\Delta z=0.1$ (dashed lines) and 0.02 per
redshift bin (dotted lines). We show cases for the flux-limited SPT
and for the same experiment with fixed $M_{\rm min}$. This figure
shows that, while missing information at $z\gtrsim 2$ can be
tolerated, having some redshift information in the region $1\lesssim
z\lesssim 2$ is very important.  While photometric redshifts in this
intermediate interval are difficult to obtain because of the lack of
prominent features, we see that relatively modest information
(redshift bias accurate to about 0.05 {\it per bin}) is sufficient to
recover most of the information obtainable with perfect-redshift
accuracy. We also checked that the error degradations in
$\Omega_M$ and $\sigma_8$ are very similar to that in $w$.

Finally, we note that our cluster count requirements were based on the
degradation in the accuracy of measuring equation of state $w$, which,
in our fiducial surveys, is measured to accuracy $\sigma_w=0.02-0.12$.
Weaker constraints on $w$, due to the inclusion of systematics, new
parameters (such as the mass-observable normalization and slope), or
else due to smaller sky coverage of the survey, will lead to {\it
weaker} redshift requirements. Nevertheless, precision measurement of
dark energy parameters is one of the principal goals of future
number-count surveys, and it is expected that they will be powerful
enough to complement concurrent supernova surveys.  Therefore,
redshift control less stringent than that advocated here would weaken
the power of number count surveys to probe dark energy.

\section{Conclusions}\label{sec:concl}

We considered how inexact redshifts affect future SNe Ia and
number-count surveys. We treated the redshifts as additional
parameters whom we assigned priors equal to their assumed measurement
accuracy. Requiring that the redshift uncertainty do not contribute more
than $\sim 10\%$ to the error budget in cosmological parameters, we
imposed requirements on the redshift accuracy.

For a future survey that studies $\sim 3000$ supernovae out to $z=1.7$
(e.g.\ the SNAP space telescope) we find that, with accurate redshift
measurements of $dz \lesssim 0.001$ for $z<0.1$ supernovae, fairly
poor redshift measurements can be tolerated at higher redshifts.
Without this accurate measurement at low redshift, however, a fairly
precise redshift measurement of $dz \lesssim 0.002$ would be required
over the full redshift range.
Photometric redshifts are probably not an option, since spectral
information is necessary to identify the SN type and control a variety
of systematic errors.  Spectroscopy can be provided using sub-pixel
interpolation of galaxy data from an on-board low-dispersion $R \sim
100$ spectrograph (which is designed to measure broad supernova
features).  Supplemental high-resolution ground-based observations
using 10m-class telescopes, adaptive optics, and OH suppression can
provide precise redshifts as necessary and to cross-check the
redshifts from the low-dispersion spectrograph.  We thus conclude that
redshift uncertainty will not significantly contribute to the error
budget in the accurate measurement of dark-energy parameters that SNAP
can deliver.

For future wide-field cluster count surveys, such as SPT, Planck or
DUET, we find that the purely statistical errors are largely
irrelevant as long as they are reasonably small (error of $\lesssim
0.02$ per cluster) because they will average out due to the large
number of clusters around any given redshift. However, the
irreducible, systematic error that doesn't decrease with increasing
number of clusters drives the redshift requirements. This irreducible
redshift-independent error has to be kept below 0.001-0.005 per
redshift bin. The widths of the redshift bins are determined by how
the redshift signature (say, the 4000\AA\ break line) goes through the
filter set of the redshift follow-up experiment, and here for
illustration we assumed filters from the Sloan Digital Sky Survey. We
found that the typical required redshift accuracy is more stringent
for X-ray surveys since they have few clusters at $z\gtrsim 1$ and
therefore use a shorter lever arm in redshift.  SZ surveys benefit
from their longer lever arm, but, of course, only if their
high-redshift clusters have decent redshift information.  Obtaining
redshifts for high-redshift clusters, therefore, should be an
important goal of any redshift follow-up survey. While the photometric
accuracy at redshifts greater than unity is highly uncertain at
present, our analysis indicates that the lack of redshift
information at $z\gtrsim 2$ does not significantly degrade the
cosmological constraints, while at redshifts $1\lesssim z\lesssim 2$
crude photometric information is sufficient to assure small
degradation in constraints on $w$ (see Fig.~\ref{fig:z_break}).
With the current rate of progress in photometric redshift techniques,
this should be a feasible goal within the next few years.

\medskip
We thank Jim Bartlett, Josh Frieman, Adrian Lee, and Tim McKay for
useful discussions. We particularly thank Eric Linder for pointing out
the importance of accurate redshifts for low-$z$ SNe, and Jim Annis and
Martin White for comments on an early draft of the paper. DH and LMK
are supported by the DOE grant to CWRU. AK was supported by the
Director, Office of Science, of the U.S.  Department of Energy under
Contract No. DE-AC03-76SF00098.

\end{document}